\documentclass[sigplan, screen, 10pt, nonacm]{acmart}
\renewcommand\footnotetextcopyrightpermission[1]{}
\RequirePackage{refcount}
\RequirePackage{soul}
\RequirePackage[alwaysadjust]{paralist}
\setul{}{1.2pt}

\makeatletter
\newcommand\extralabel[2]{{\edef\@currentlabel{\@currentlabel#2}\label{#1}}}
\makeatother

\newcommand\pcode[1]{\textsf{#1}}

\begin{document}

\title{Safe Sharing of Fast Kernel-Bypass I/O Among Nontrusting Applications}

\author{Alan Beadle}
\email{beadle@alfred.edu}
\orcid{0000-0003-1243-8213}
\affiliation{%
  \institution{Alfred University}
  \city{Alfred}
  \state{NY}
  \country{USA}
}

\author{Michael L. Scott}
\email{scott@cs.rochester.edu}
\orcid{0000-0001-8652-7644}
\affiliation{%
  \institution{University of Rochester}
  \city{Rochester}
  \state{NY}
  \country{USA}
}

\author{John Criswell}
\email{criswell@cs.rochester.edu}
\orcid{0000-0003-2176-3659}
\affiliation{%
  \institution{University of Rochester}
  \city{Rochester}
  \state{NY}
  \country{USA}
}

\renewcommand{\shortauthors}{Beadle, Scott, and Criswell}

\begin{abstract}
Protected user-level libraries have been proposed as a way to allow
mutually distrusting applications to safely share kernel-bypass
services.  In this paper, we identify and solve several previously
unaddressed obstacles to realizing this design and identify several
optimization opportunities.  First, to preserve the kernel's ability to
reclaim failed processes, protected library functions must complete in
modest, bounded time.  We show how to move unbounded waits outside the
library itself, enabling synchronous interaction among processes without
the need for polling.  Second, we show how the bounded time requirement
can be leveraged to achieve lower and more stable latency for
inter-process interactions.  Third, we observe that prior work on
protected libraries is vulnerable to a buffer unmapping attack; we
prevent this attack by preventing applications from removing pages that
they share with the protected library.  Fourth, we show how a trusted
daemon can respond to asynchronous events and dynamically divide work
with application threads in a protected library.

By extending and improving the protected library model, our work
provides a new way to structure OS services, combining the advantages of
kernel bypass and microkernels.  We present a set of safety and
performance guidelines for developers of protected libraries, and a set
of recommendations for developers of future protected library operating
systems.  We demonstrate the convenience and performance of our approach
with a prototype version of the DDS communication service.  To the best
of our knowledge, this prototype represents the first successful sharing
of a kernel-bypass NIC among mutually untrusting applications.  Relative
to the commercial FastDDS implementation, we achieve approximately 50\%
lower latency and up to 7x throughput, with lower CPU utilization.
\end{abstract}

\settopmatter{printfolios=true, printacmref=false, printccs=false}
\maketitle

\section{Introduction}

In most operating systems, processes rely on the OS kernel to manage
resource sharing and isolation.  Processes interact with the kernel
through system calls, which entail expensive mode switches but protect
the integrity of shared resources while supporting interprocess
interaction.

\emph{Kernel bypass} techniques aim to mitigate switching overhead by
granting a process direct (typically exclusive) access to some normally
shared resource.  Example systems include Intel's DPDK
(networking)~\cite{dpdk} and SPDK (storage)~\cite{spdk} libraries.
While the performance gain can be substantial, such libraries lack the
isolation on which many applications rely for security and stability:
the absence of a trusted mediator means that bypass-based systems cannot
tolerate malicious or buggy applications.

An opposite trend in operating systems research is to sacrifice
performance for isolation.  \emph{Microkernels}, in particular,
prioritize modularity, fault containment, and incremental development
and deployment by placing separate OS services in separate address
spaces.  Applications then access these services via interprocess
communication (IPC) mechanisms provided by the remaining, minimal
kernel.  Unfortunately, microkernels tend to suffer from substantial
communication overhead~\cite{hartig-sosp-1997}, limiting their adoption.

More recently, the ERIM~\cite{erim-usenix-sec-1029} and
Hodor~\cite{hedayati2019} projects have proposed to use memory
protection keys (MPK---marketed by Intel as Protection Keys for
Userspace, or PKU~\cite{intelsdm}), to implement lightweight protection
domains within an address space.  This strategy suggests the possibility
of a \emph{protected library} that can be linked simultaneously into
mutually untrusting applications, allowing memory and memory-mapped
devices to be safely shared in user space.  The Hodor team
used protected libraries to build a (single-process) instance of the
Silo in-memory database~\cite{tu-sosp13-silo} and a version of the Redis
NoSQL server~\cite{redis} that communicated with remote clients via
Intel's DPDK kernel-bypass network stack~\cite{dpdk}.  A separate,
follow-on project~\cite{kjellqvist2020} developed a protected-library
version of the memcached key-value store~\cite{libmemcached}.

Unfortunately, the protected library examples explored to date have
failed to expose---much less solve---the full set of problems associated
with sharing memory-mapped resources safely in user space.  In
particular, all have entailed library calls that are essentially
self-contained: they never handle asynchronous events, condition
synchronization, or synchronous interaction among applications.  Our
work overcomes these limitations, significantly expanding the types of
kernel services that can safely be moved into protected libraries.
Among other things, we direct asynchronous events to a trusted daemon
and enable low-latency, synchronous communication among independent
processes while still respecting the programming constraints required
for isolation and resource integrity---in particular, the requirement
that library calls complete in modest bounded time.

Our contributions allow protected libraries, for the first time, to
combine the advantages of microkernel and kernel-bypass techniques.
Specifically:
\begin{enumerate}
\item
  We show how to move unbounded waits outside of the protected library
  without the need for polling, by leveraging shared (but protected)
  futex mappings (Sec.~\ref{sec:suspending}).
\item
  We show how the bounded time requirement can be leveraged to
  parallelize thread wakeup with other work, yielding lower and more
  stable latency (Sec.~\ref{sec:eagernotification}).
\item
  We introduce designated buffers through which to pass data into a
  protected library, avoiding a memory unmapping vulnerability in prior
  work (Sec.~\ref{sec:permbuf}).
\item
  We introduce the notion of a trusted daemon to respond to asynchronous
  events and study the ways it can divide work with other threads
  (Sec.~\ref{sec:async}).
\item
  We present a set of guidelines for the developers of protected
  libraries, both to maintain safety and to maximize performance
  (Sec.~\ref{sec:guidelines}).
\item
  We enumerate recommendations for future protected library operating
  systems, to solve otherwise challenging safety issues without
  additional overhead (Sec.~\ref{sec:features}).
\item
  As validation of our contributions, we present a protected library
  version of the DDS communication service
  (Sec.~\ref{sec:implementation}).
\end{enumerate}

Our DDS implementation demonstrates safe, efficient, synchronous
interaction among mutually untrusting processes, with a shared
memory-mapped device---in our case, the network interface.  Leveraging
this sharing, our prototype achieves significantly higher and more
consistent performance than the industry-standard FastDDS
system~\cite{fastdds} (Sec.~\ref{sec:performance}).

\section{Background}

\subsection{Hodor}
\label{sec:hodor}

Hodor~\cite{hedayati2019} is a system designed to support user-space
protected libraries.  A protected library is similar to a shared
library, but with the option to control access to both code and data
depending on whether execution is currently in the library or in the
main application.  By granting library code access to a shared memory
region that is otherwise inaccessible, Hodor allows both memory and
peripheral devices to be shared safely among library instances in
mutually distrusting applications.  The author of a protected library
can, with care, limit the operations that can be performed on the shared
resource, thereby maintaining correctness and security invariants.
Critically, Hodor permits calls into a protected library without leaving
unprivileged execution mode.  As a result, calling a protected library
function is faster than an equivalent system call.

Under the Hodor threat model, the hardware and operating system are part
of the trusted computing base.  Each protected library is trusted by
applications that link to it, but need not be trusted by the operating
system.  Applications are untrusted and assumed to be compromised: they
may make arbitrary calls into the protected library via the exposed API
but are prevented by Hodor from interacting with it in any other way.
Internally, the library is responsible for resource management (e.g., of
communication buffers) and for any other constraints on the correctness
of the service it provides (e.g., regarding information flow).

The code and data of a protected library comprise one or more protected
memory regions.  Like a conventional shared library, code is compiled to
be position-independent; data is accessed through address space-specific
indirection tables.  Hodor ensures that only trusted code can directly
read and write a library's protected memory regions.  If the operating
system trusts the library to manage physical devices, these may also be
included in the library's memory space.  By default, a thread executing
in a protected library can access arbitrary memory in the address space
of the current process, but this can be restricted if desired.  Special
\emph{trampoline} code, accessible to both the application and the
library, serves to transition between application and library execution.

The preferred implementation of Hodor comprises a modified Linux 5.4
kernel and a trusted loader that employs Intel's Memory Protection Keys
(MPK), also known as Protection Keys for Userspace
(PKU)~\cite{intelsdm}.  MPK allows an application to restrict and grant
access to memory pages on a per-thread basis (subject to maximum
permissions embodied in the page table) without making system calls.
Page table entries contain a new 4-bit key which is used to group pages
in up to 16 protection groups.  A special \emph{PKRU} register can then
be set to restrict read and/or write access to the pages of each of the
groups.  PKRU contents can be changed at any time using the
(unprivileged) WRPKRU instruction.  On every user-level memory access,
the processor determines the minimum of the page table rights and the
rights indicated (for the relevant protection group) in the PKRU.  An
access exceeding this minimum violation results in a segmentation fault.
Several groups have explored methods to mitigate the limitation of 16
protection groups~\cite{gu2022, park2019}.

When a Hodor application calls a protected library function, the PKRU
must be changed to give the library function access to the library's
data.  The change is performed by \emph{trampoline} code that includes a
WRPKRU instruction.  To ensure that such instructions occur \emph{only}
in trampoline code (so as to prevent applications from bypassing Hodor's
protections), Hodor's trusted loader finds any other WRPKRU instructions
and uses hardware watchpoints to prevent their execution.

Hodor delays signal delivery to any thread executing within a protected
library to prevent leveraging signal handlers to bypass protection.  To
ensure the consistency of library data, the kernel also declines to
terminate a process while any of its threads are executing in the
library.  Instead, it halts all other threads, sets watchpoints to catch
the library threads when they return to the application, and terminates
the process when all such threads have done so.  To preserve the
kernel's ability to reclaim failed processes all protected library calls
are required to complete in modest bounded time.  If the time limit is
exceeded, the library itself is assumed to be buggy and all applications
using it are terminated.%
\footnote{Sadly, the current, experimental version of Hodor does not
    prevent all interrupts, and a process may be preempted while in a
    critical section within the protected library, potentially resulting in
    priority inversion.  Sec.~\ref{sec:configuration} describes how we
    avoided this problem in our experiments.}

Hodor has been used to implement single-process instances of the Redis
in-memory database and the DPDK kernel-bypass network
stack~\cite{hedayati2019} and to build a version of the memcached key
value store that uses protected library calls instead of message
passing~\cite{kjellqvist2020}.  The memcached work demonstrated that
mutually distrusting applications can share a protected library safely,
with direct, function-call access that is dramatically faster than
IPC\@.  Process interactions in that work, however, were entirely
one-sided: no application ever needed to interact synchronously with
another.

Synchronous interactions introduce a new set of questions.  First, how
does an application wait for activity on the part of an untrusted peer?
How does the peer alert it when it can continue?  How is waiting even
possible in a system like Hodor, where library calls are required to
complete in bounded time?  Second, is the requirement for bounded-time
operations always a negative, or can we sometimes exploit its guarantee
of progress?  Third, when buffers are passed between an application and
a protected library---or between applications \emph{through} a protected
library---how do we minimize copying?  How do we ensure that no
application activity (e.g., unmapping of buffers, performed in another
thread) can compromise the safety of a protected library call?  Fourth,
if a library manages a memory-mapped device that is to be shared among
applications, how do we handle asynchronous events?  What application
should field them?  We address these questions in
Secs.~\ref{sec:suspending} through~\ref{sec:async}.

\subsection{DDS}

As a concrete example of a kernel-bypass service with synchronous
process interactions, memory-mapped devices, and asynchronous events, we
consider the Data Distribution Service~\cite{omg-dds}.  DDS is a
specification for publish-subscribe middleware for distributed real-time
data exchange.  Commercial and noncommercial DDS implementations are
widely used in aerospace, robotics, and other fields; Among other
things, they comprise the primary communication layer in
ROS\,2~\cite{rosdds}.  The specification includes distributed discovery
of peers and fine-grain quality of service (QoS) control.

In DDS, a \emph{topic} represents a communication channel for sending
data from publishers to subscribers.  A topic has a name, a type, and
various QoS settings.  Data is published to a topic in the form of
messages called \emph{samples}, which must be of the type described by
the topic. We will occasionally use the term \emph{messages} to refer to
DDS samples.

One of the main design priorities of DDS is simplicity of use.
Application programmers do not need to explicitly search for peers, nor
marshal messages.  When one application creates a topic on which to
publish, peers subscribing to that topic by name will receive data
published to it, regardless of their location.  A subscriber can poll a
given topic until a new sample is available.  Alternatively, it can
suspend until some condition is met---for example, until a new message
arrives on one of some set of specified topics.

DDS supports QoS settings on a per-topic basis.  Several of these
settings address \emph{reliability}.  Topics can be configured with
\emph{best effort} reliability, meaning that message loss can be
tolerated.  Alternatively, topics can be configured to guarantee
reliable delivery of all messages.  Independently, they can be
configured to keep some predefined number of recent messages or to keep
each message until it has been received by each subscriber.  Other QoS
settings address \emph{durability}.  If a subscriber was not known at
the time that a sample was sent, but is discovered later, these settings
can be used to deliver recent samples to the new subscriber.

Given its emphasis on real-time IPC, DDS requires synchronous
interactions among processes---something previous work on protected
libraries has not addressed.  With its emphasis on distribution, it also
has the potential to benefit greatly from memory-mapped access to the
network interface (NIC), provided that we can share that interface
safely and harvest interrupts from it.  The relatively high level of DDS
semantics suggests an opportunity to access complex functionality and
data with fewer API calls (and therefore lower overhead) than would be
needed in a simpler (e.g., socket-based) system.  Moreover, this
complexity seems like a poor fit for integration into the OS kernel,
which needs to remain general-purpose.%
\footnote{Twenty years ago, there appears to have been at least one
    proposal for kernel integration (\url{www.navysbir.com/05_3/6.htm})
    but the idea never caught traction.}

\subsection{Related Work}

In the context of a traditional monolithic kernel, kernel bypass
techniques allow authorized applications to access certain hardware
devices directly.  This approach preserves familiar abstractions while
selectively allowing some operations to be performed with very little
overhead.  DPDK~\cite{dpdk}, for example, allows applications to perform
networking operations in user mode.  SPDK~\cite{spdk,
  spdk-cloudcom-2017} provides similar functionality for storage
devices.

Unfortunately, most kernel bypass work requires that mutually
distrusting applications never access the same device---otherwise, one
might monopolize the shared resource, corrupt the data of the other
application, or steal confidential information.  DPDK and SPDK assume
that any applications with shared access are mutually trusting.  Our
work uses the Hodor kernel, which is designed to support safer kernel
bypass via protected libraries.  Hodor permits us to bypass the kernel
when accessing shared resources, but preserves the ability to enforce
controls on how the resources are used.

DerechoDDS~\cite{rosa2021a, rosa2021b} transports data over RDMA in user
mode to achieve higher networked DDS performance.  However, this
approach fails to isolate mutually distrusting applications on the same
machine.  More recent work~\cite{bode-percom-2024, bode-percom-2025} has
shown that existing DDS implementations can be retrofitted to use DPDK
kernel bypass (and to a lesser degree with XDP~\cite{xdp-2018}, an
in-kernel packet processing system) for significantly improved
performance.  The authors acknowledge, however, that there is currently
no way to maintain isolation between processes with these techniques, or
even to protect against accidental memory accesses.  Our work achieves
similar performance improvements and prevents both intentional and
accidental errors.  Given the unique challenges of protected libraries,
we opted to develop a DDS prototype from scratch.  In doing so, we have
developed guidelines (Sec.~\ref{sec:guidelines}) that may facilitate
retrofits of existing software in the future.

Snap~\cite{marty-sosp-2019} preserves a monolithic kernel while adding a
trusted user-level server process to manage shared resources.
Applications invoke operations on a resource via request and response
buffers located in memory shared with the server.  Like Hodor, Snap
moves resource management out of the kernel while preserving process
isolation, and facilitates the development of custom interfaces to meet
application needs. Because the server runs as a separate process,
however, invoking a single operation is guaranteed to require either
cross-core communication between processes or an expensive context
switch to the other process and back. In contrast, Hodor uses the
calling thread's own context to execute operations.  Running in user
mode, the Snap daemon must poll frequently to ensure low latency.  This
strategy can waste CPU time when events are less frequent.  We discuss
our solution to this inefficiency in Sec.~\ref{sec:async}.

Xia et al.\,\cite{xia2022} prototype an IPC mechanism that bypasses the
kernel.  Their design requires special CPU instructions so cannot be
employed on current commodity hardware.  We believe their hardware could
be used to implement protected libraries if it were deployed.

\section{Protected Library System Design}
\label{sec:design}

\begin{figure*}
\begin{minipage}[b]{0.41\textwidth}
\Description{Independent applications passing messages through the
  kernel using FastDDS.}
\caption{Conventional IPC. \textmd{The address space of application A
    is shown in red.  It includes a heavyweight protection boundary
    between the app and the kernel.  All communication passes through
    the kernel.  Trusted code (here only in the kernel) is shown in
    blue.  Remote communication requires execution in four contexts.}}
\label{fig:fastdds}
\end{minipage}
\hfill
\includegraphics[scale=0.68]{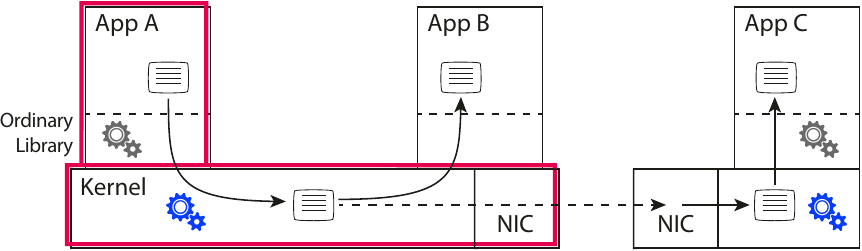}
\begin{minipage}[b]{0.35\textwidth}
\Description{Applications communicating through a protected library.}
\caption{Protected Library IPC. \textmd{Applications on the same machine
    share a single library instance, allowing them to communicate safely
    in user space.  The address space of each application includes the
    data buffers of all applications, together with the network
    interface.  There is a lightweight protection boundary between the
    application and the library.  Library code is trusted, and can
    safely access shared metadata, even when executed by application
    threads.  Remote communication requires execution in only two
    contexts.}}
\label{fig:hodordds}
\end{minipage}
\hfill
\raisebox{2ex}{\includegraphics[scale=0.68]{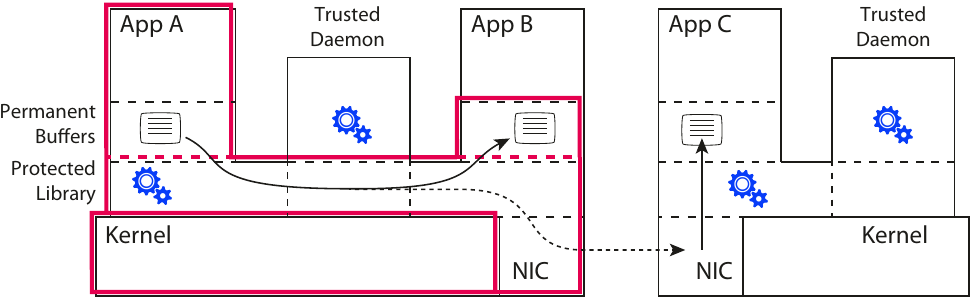}}
\end{figure*}

Our protected library system builds on the original Hodor
design~\cite{hedayati2019}.  As illustrated in Fig.~\ref{fig:fastdds},
using IPC as an example, conventional library communication entails
context switches among three separate address spaces---the sender, the
kernel, and the receiver.  For communication over the network, four
contexts are involved.  In contrast, protected libraries
(Fig.~\ref{fig:hodordds}) allow the sender's thread to access a local
receiver's buffers directly and safely, with no context switch required.
Remote communication involves only one context on each machine.

To support the requirement that library calls complete in modest,
bounded time, we assume (as implied in the original Hodor
paper~\cite{hedayati2019} but not actually implemented in the
open-source code base) that the kernel will temporarily avoid preempting
threads that are executing in a protected library.  This functionality
can be implemented in a manner reminiscent of the \pcode{schedctl}
mechanism in the Solaris kernel \cite{dice-schedctl} and of related
mechanisms in several earlier academic systems \cite{edler-uwus-1988,
  marsh-sosp-1991, anderson-tocs-1992}.  If a library call runs past the
normal end of a thread's time quantum, the kernel can deduct the extra
time from the start of a subsequent quantum, for the sake of fairness.
A call that takes more than a published maximum time bound can be
assumed to indicate a buggy library, users of which can be terminated
with prejudice.

To avoid exceeding the kernel's time bound, a protected library call can
never wait for anything (e.g., the arrival of a network packet) that
isn't guaranteed to happen soon.  It cannot use blocking synchronization
of any kind, nor can it use spin locks to protect more than the shortest
critical sections.  In a similar vein, it can never return to the
application (making it vulnerable to indefinite delay or to termination
due to user bugs) while holding a library lock.  To lessen the burden of
these restrictions, we show (in Sec.~\ref{sec:suspending}) how to safely
move unbounded waits outside the protected library.  In
Sec.~\ref{sec:eagernotification} we show how to \emph{leverage} the
bounded-time requirement to improve performance.

\label{sec:memory}
As in the original Hodor paper, we maintain protected library state in
an MPK domain that is inaccessible to application code. This domain
includes a \emph{protected heap} for shared data.  The heap is available
to all processes using the same protected library instance but can only
be accessed by library functions---not by untrusted application code.
Objects allocated in the protected heap can be assigned an owning
process and a unique ID, reminiscent of a file descriptor.  Library
calls can then refer to resources or invoke methods in terms of these
descriptors.  While running in the protected library, threads may access
resources belonging to a peer, but the library should ensure that they
never attempt to invoke a method of an inappropriate object.

\label{sec:getpid}
To determine what constitutes ``appropriate,'' a protected library needs
a way to verify the identity of the current process.  It could, of
course, use the \pcode{getpid(\,)} system call, but given that much of
the rationale for protected libraries is to avoid the overhead of system
calls, we assume instead that the kernel records the current PID in a
location visible to protected library code.  Such a mechanism could
easily be provided using Linux's vDSO~\cite{vdso}.

To ensure that resources cannot be accessed (maliciously or
accidentally) after the owning thread terminates, the kernel must guard
against PID reuse attacks.  One approach would be to select new PIDs
randomly from a large set of possible values; with sufficient entropy,
attackers would be unable to engineer reuse.  If PIDs were sufficiently
large (64 bits, say), we could use monotonically increasing values and
never worry about rollover.  Alternatively, the kernel could arrange to
perform an upcall into the protected library (handled, perhaps by a
trusted daemon---Sec.~\ref{sec:async}) whenever a client process
terminates; the library could then account for this in its PID access
control checks.

Registers can be used to pass small amounts of data between applications
and protected libraries, but larger amounts require shared buffers for
efficiency.  To minimize copy operations and facilitate work sharing,
the buffers of \emph{all} processes should ideally be visible to all
threads running in the library.  It is essential however, that the
library validate pointers to these buffers before use, and that
applications be prevented from making them invalid (e.g., by unmapping
the buffers) while they are in use.  We discuss this issue further in
Sec.~\ref{sec:permbuf}.

\subsection{Safe Thread Suspension}
\label{sec:suspending}

As noted in Sec.~\ref{sec:hodor}, a Hodor library function must never
suspend or spin for an unbounded length of time.  This means that if a
thread must wait for some condition to be met, it cannot do so within
the protected library unless the duration of the wait is brief and known
in advance.  In principle, one could imagine calling into the protected
library in every iteration of a spin loop, but trampoline crossing costs
make such a strategy highly undesirable.

Instead, we arrange for threads to safely suspend \emph{outside} the
protected library and to be awoken by threads \emph{inside} the library.
To achieve this, we leverage the shared mapping of the protected heap
and the ability to surround a protected library call with an unprotected
wrapper.  If the protected call determines that the thread must wait, it
returns an indication of this to the wrapper, which then suspends
outside the library.  Some other thread in the protected library
(perhaps from another application) will awaken the suspended thread when
the prescribed condition is met, at which point the wrapper routine can
call into the library again.

Our design requires that an arbitrary library thread be able to wake an
arbitrary peer.  To do so, we leverage the fast userspace mutex
(\emph{futex})~\cite{franke-ols-2002} mechanism.  Futexes require that a
memory word be mapped into the address spaces of both the sleeper and
the thread that wakes it.  Ours are located in protected memory and are
therefore available to all its clients, albeit protected from direct
access by application code.  From within the library, the futex is
accessed in the usual way, by threads that make conditions true.  When
protected code determines that a thread must wait, it returns the
address of the futex word to the surrounding wrapper routine, which then
performs a \pcode{FUTEX\_WAIT} syscall \emph{outside} the protected
library.  This works as the wrapper routine doesn't read or write the
futex word; it merely passes its address to the OS kernel.  When the
thread is awoken by a peer, it calls back into the protected library for
its next (time-bounded) step.

\subsection{Leveraging Bounded Time Completion}
\label{sec:eagernotification}

While it is tempting to assume that bounded execution time is merely a
burden to the authors of protected library functions, we argue that it
can in fact enable useful optimizations.  Consider a scenario in which
one thread must wait for another thread to complete some task---e.g., to
fill a buffer it wants to read.  If peer threads may fail independently,
as is normally the case when they belong to separate processes, a
failure in one thread could cause the other to wait indefinitely.  One
possible solution would be to notify the waiting thread only when the
buffer is completely filled, at which point a failure of the producer
thread would not jeopardize the progress of the waiting thread.
Timeouts could also be used to abort the wait if the operation takes
longer than expected.  In both of these solutions, uncertainty about
when the copying thread will finish can lead to wasted CPU time.

If the producer thread is running in a protected library, however, we
know that the operation will take bounded time.  If the copying code has
no failure path, the waiting thread can safely be notified (e.g., via
futex) as soon as copying begins.  That is, the bounded time requirement
enables \emph{eager notification}.  As long as the waiting thread can
tell which parts of the buffer have been filled (and can spin if it gets
ahead of the producer), we can parallelize the producer's copy step with
the preparation (e.g., wakeup) of the waiting thread, with no risk of
having to abort after a timeout.  Used carefully, this technique can
reduce both the magnitude and the variance of synchronous operation
latency.  In order to benefit, the producer thread must have a
nontrivial amount of useful work to perform after waking the consumer
thread.  The duration of that work should approach the time required for
a thread to wake from sleep.  Sec.~\ref{sec:hodordds_EN} describes how
we use this technique in HodorDDS\@.

\subsection{Defense Against Unmapping Attacks}
\label{sec:permbuf}

Just as an OS system call must validate pointers passed from user
programs, so too must a Hodor library validate pointers passed to its
routines.  First, it must ensure that each pointer falls within the
memory shared between the application and protected library.  Otherwise,
an application could trick a protected library into leaking or
corrupting its own internal data.  Second, it must ensure that the
memory to which the pointer refers is mapped with the correct
permissions.  An application might accidentally pass a pointer to
unmapped memory.  Alternatively, an application might try to use {\tt
  mmap()} and {\tt mprotect()} in one thread to unmap or change the
permissions of a portion of the shared heap while another thread is
calling a protected library routine.

Solutions for OS kernels do not work for protected libraries.  To
validate pointers, the Linux kernel maps its memory into the upper
portion of each process's address space; it validates pointer arguments
by checking that the buffer to read or write is located below a
hard-coded address~\cite{LinuxKernelBook2}.  To handle application
buffers within unmapped, unreadable, or unwriteable memory, the OS
kernel uses special data copying functions (e.g., {\tt
  copy\_from\_user()}) to copy data in application buffers into the
kernel before use; the OS kernel's trap handlers detect when these
copying routines cause a trap and jump to recovery
routines~\cite{LinuxKernelBook2}.  Since protected library code executes
in user mode, it cannot use hardware trap handlers to recover from
memory protection faults when reading and writing buffers passed from
the application.

Applications communicating via shared memory face similar safety
challenges, but the solutions used in such cases are also insufficient
for protected libraries.  Processes can only affect their own memory
mappings, meaning that if the process verifies that a pointer points to
a mapped address, it is safe to dereference.  If the process unmaps its
own buffer while dereferencing the validated pointer, only the faulty
process is crashed.  In a protected library however, a process may unmap
its own buffer while another of its threads is accessing the buffer from
within the library, potentially leaving the library in an inconsistent
and unusable state.

To address these challenges, our design enhances the Hodor kernel.
First, we extend \pcode{mmap(\,)} with a flag that permits a protected
library to map memory that both the library and the application can read
and write, but that can be unmapped (or have its permissions changed)
only by the library.  This memory can then be used to pass data to and
from protected library functions.  As this memory is permanently mapped
(but not pinned; the OS kernel can swap it out as needed), and library
threads (from any process) can read and write the buffers of their
peers.

Since protected libraries map these \emph{permanent buffer} regions,
they will always know their locations.  Most use cases would only need
one region with which to transfer data between the library and a given
application, so validating that pointers passed from the application are
within this region can be done with two comparison instructions.  For
more complex scenarios, multiple regions might need to be mapped, but
pointers could still be validated against a set of regions in
logarithmic time.

For applications that use several protected libraries, we may need to
limit how much address space the protected libraries can collectively
reserve (this is especially true if the libraries are mutually
distrusting).  To do this, we propose adding a new per-process resource
limit for how much address space can be reserved with our new
\pcode{mmap(\,)} feature.  Increasing the limit will require privilege;
the Hodor kernel can add a new Linux capability for increasing this
limit.

When either an application or library thread needs to pass information
across the trampoline boundary, it first allocates a portion of the
permanent buffer of the relevant process.  We used a simple memory
allocator that divides the buffer into blocks and tracks whether each
block is owned by the library or the application.  The allocator code is
duplicated in the application wrappers and in the protected library.
After a block is filled with data, its ownership status is changed to
the recipient (application or library) and a pointer to it is passed via
register.  Care must be taken to ensure that protected library functions
are never forced to block due to unavailability of permanent buffer
space, even if a badly designed or malicious application corrupts the
allocator metadata, which is itself contained in the permanent buffer.

In HodorDDS, permanent buffer mappings allow threads, when running in
the library, to directly read buffers belonging to their peers for
faster delivery.  (see Fig.~\ref{fig:hodordds}).  In principle, we could
also use permanent buffers to communicate between library threads and
application threads without the need for trampoline crossings.  This
approach would offer only minimal benefits in HodorDDS, but there may be
cases where it would be advantageous.  We believe it is more likely to
be useful for receiving data from the library than for sending data into
the library because there is no way for the application to immediately
notify threads inside of the library that data is ready to be sent,
whereas the futex mechanism of Sec.~\ref{sec:suspending} enables low
latency notification of application threads by library threads.

\subsection{Fielding Asynchronous Events}
\label{sec:async}

Many OS services need to respond to asynchronous events in a timely
manner.  Allowing applications to perform tasks with their own threads
in user mode has performance advantages but does not guarantee that
events will be detected within strict time bounds.  In particular,
services that allow applications to share and interact with I/O hardware
often rely on interrupt handlers running in kernel mode to allow the
kernel to respond to events.  Because protected libraries operate in
user mode, such handlers cannot be used without modification.  A common
alternative strategy in kernel bypass systems is to poll devices
frequently enough to keep average latency low.  Such polling can result
in a shorter code path, but can also waste cycles when events are
infrequent.

To address this tradeoff---and to enable the interrupt option for shared
kernel-bypass I/O---we introduce the notion of a \emph{trusted daemon}.
The daemon is considered part of the trusted computing base and should
be run whenever its protected library is loaded.  Because the daemon is
trusted, it is able to bypass Hodor trampolines used by untrusted
applications to enter the library, meaning that a given library function
incurs less overhead when called by the daemon than it does when called
by an application thread.  When running in the protected library, the
daemon has the same memory access rights as any of the application
processes, meaning that it can also read and write the protected heap
and the application data buffers described in Sec.~\ref{sec:permbuf}.

The use of a trusted daemon enables a hybrid approach that achieves low
CPU utilization and low latency not only when events are infrequent but
also when events are so frequent that the interrupt code path is too
long to service each event on time.  When a peripheral device is
configured for kernel bypass, the kernel can still receive interrupts
from that device.  By arranging for the kernel to wake a suspended
daemon when such an interrupt occurs, protected libraries can respond to
events promptly without constant polling.  Note, of course, that the
daemon is not receiving interrupts directly: it must suspend (or attempt
to suspend) to detect that the kernel has received an interrupt.

As noted above, the code path for pure user mode polling is shorter than
that of the interrupt-driven method.  If the event rate is sufficiently
high, then the interrupt-driven approach leaves insufficient time to
process events, which causes the throughput to plateau.  By detecting
when this occurs, protected libraries can opportunistically switch
between interrupt and polling modes to achieve the best possible
performance.  We explore this topic further in
Sec.~\ref{sec:dds-daemon}.

\section{Guidelines for Future Developers}

\subsection{For Developers of Protected Libraries}
\label{sec:guidelines}

A protected library must be written with care if it is to function both
safely and efficiently.  We summarize guidelines here. Items
\ref{item-bounded} through \ref{item-cleanup} pertain to safety, and the
remaining items pertain to maximizing performance.

\newcounter{enumcleanup}
\begin{enumerate}
\item
    A protected library must never do anything that may take more than
    the bounded time guaranteed to be allotted by the kernel.
    Particularly time-consuming operations may need to be broken into
    pieces.  More important, a library call must never spin or block
    while waiting for activity (in a device or another thread) that is
    not guaranteed (absent catastrophic failures) to occur in bounded
    time.  To get around this limitation, a library call may return a
    futex on which a thread can wait \emph{outside} the library.
    Certain system calls may be acceptable in protected library
    routines, if the kernel guarantees they will be brief.  Bounded
    spins may also be acceptable, if the library can do something
    reasonable if the spin times out.
\label{item-bounded}
\item
    A protected library must carefully check the validity of all
    arguments passed to it by applications, just as a syscall must.  As
    Sec.~\ref{sec:permbuf} notes, a protected library should use our
    suggested Hodor kernel enhancements to allocate permanent buffers,
    and it should check that all input pointer arguments point into this
    buffer space.
\label{item-argcheck}
\item
    Resources within a protected library, associated with a given
    application, must be reclaimed when the application exits.  In the
    usual case, an application should invoke a cleanup routine
    immediately prior to completion.  To accommodate abnormal
    termination, the library may perform periodic memory reclamation.
    To assist with this collection, the kernel should provide a way to
    determine whether a given application is still running.
    Alternatively, a production version of a system like Hodor might
    provide an upcall~\cite{anderson-tocs-1992} to a library when a
    client exits.
\label{item-cleanup}
\item
    To minimize context switches, a protected library should, whenever
    possible, perform bookkeeping operations using application threads
    that have already called into the library.  Dedicated (daemon)
    threads should be created only when some operation needs to occur on
    a more regular (e.g., time-driven) basis than application calls.
\label{item-appthread}
\item
    Given that trampoline calls, while faster than syscalls, are still
    significantly more expensive than ordinary subroutine calls, a
    protected library should do as much work as possible in each library
    call, without violating the bounded time requirement (item 1).  That
    is, the reduced call overhead relative to a system call is not
    sufficient to warrant the use of finer grained operations than
    system calls normally perform. On the contrary, the ability to
    develop custom operations in a protected library provides the
    opportunity to replace sequences of system-call-based primitives
    with fewer custom library calls, further minimizing the relative
    overhead per operation.
\label{item-api}
\item
    As the introduction to Sec.~\ref{sec:getpid} explains, a production
    system should provide a fast mechanism by which the library can
    reliably identify the currently running thread.  A protected library
    should use this mechanism over existing kernel system calls to
    maximize performance.
\end{enumerate}

Some of these items---\ref{item-argcheck} and \ref{item-cleanup}, for
example---resemble challenges faced in operations implemented in the
kernel and invoked with syscalls (although the solutions are necessarily
different).  Item \ref{item-bounded} is an issue for traditional signal
and interrupt handlers.  Items \ref{item-appthread} and \ref{item-api}
can be considered new opportunities that arise with protected libraries.

\subsection{For Developers of Protected Library OSes}
\label{sec:features}

Summarizing observations from Sec.~\ref{sec:design}, a production
version of the Hodor kernel should include
\begin{enumerate}
\item
    fast, unspoofable identification of the running thread.
\item
    avoidance of preemption during protected library calls.
\item
    buffers that can be unmapped from an application only from within a
    protected library (Section \ref{sec:permbuf}).
\item
    a mechanism by which a protected library can determine (or be
    alerted) that a client application has terminated (Section
    \ref{sec:guidelines}, item~\ref{item-cleanup}).
\end{enumerate}

\section{Prototype Implementation}
\label{sec:implementation}

As a validation of our ideas, we have prototyped a protected library
version of DDS, which we call HodorDDS.  Our prototype does not rely on
any preexisting IPC libraries and comprises approximately 5,000 lines of
C++~code.  It runs on a modified Linux kernel with support for Hodor
protected libraries.  To use a newer version of the Linux kernel, we
took the changes that the original Hodor authors made to the Linux
kernel~\cite{hedayati2019} and merged them into the Linux 5.4.0 kernel.
We use DPDK to send and receive network packets from user mode and
implement a subset of the RTPS~\cite{omg-rtps} network protocol used by
most other DDS implementations.

As an experimental prototype rather than a production system, our
implementation supports only a subset of the DDS
specification~\cite{omg-dds}---a subset chosen to expose all the
fundamental challenges faced by the protected library approach to OS
services, including management of shared networking hardware.  Extension
to the full API would be tedious but straightforward.

\subsection{API and Memory Regions}

The HodorDDS protected heap contains structures that correspond to
objects described in the DDS specification, including participants,
topics, writers, readers, publishers, subscribers, waitsets, and
samples.  When these objects are created, they are considered to belong
to the process that created them.  Operations on these objects are
vetted as described in Sec.~\ref{sec:memory}.  DPDK also includes a
dedicated heap and allocator for buffers to contain inbound or outbound
packet data (mbufs).  We do not permit applications to directly read or
write these mbufs since improper access could expose vulnerabilities in
DPDK or result in corrupted packets being sent.

Applications send messages (samples) using \pcode{hdds\_write(\,)} and
receive messages using \pcode{hdds\_take(\,)}.%
\footnote{For brevity, we discuss only the minimal subset of the DDS API
    employed in our benchmarks.}
The former takes a descriptor for a valid DDS writer object and a
pointer to a prepared sample as parameters; the latter takes a
descriptor for a valid DDS \emph{reader} object, a pointer to a buffer
into which messages should be copied, a pointer to a buffer used to pass
metadata about each sample returned, and specifications of the capacity
of the sample buffer and the maximum number of samples to be returned.
Return values from \pcode{hdds\_take(\,)} indicate the number of
messages received.

In our implementation, \pcode{hdds\_take(\,)} is actually an external
wrapper that checks whether the permanent buffers still contain any
unread samples before calling into the protected library.  Samples can
be returned in batches even if the application only requests one sample
at a time.  (In our tests this was an uncommon case, as each sample was
usually handled before the next one arrived.)

Our Hodor Linux kernel lacks the vDSO support for fast access to the PID
and the PID reuse mitigations described in Sec.~\ref{sec:getpid}.  Our
HodorDDS prototype performs a dummy ``validation'' step that mimics the
(minuscule) cost of checking a PID made visible by the kernel.

Our prototype supports transferring large samples over the network via
fragmentation.  The permanent buffers (Sec.~\ref{sec:permbuf}) are
specifically not used for reassembly of fragmented DDS samples, because
there may be many recipients for the same sample and each buffer can
only be accessed from outside the library by the owning process.
Instead, fragmented samples are reassembled in a buffer allocated from
the protected heap. Regardless of whether a received sample originated
locally or on a network peer, and regardless of whether or not the
sample was fragmented during network transport, receiving threads must
copy the completed sample into their own permanently mapped buffers from
within the library. In principle, this step could be done by one thread
on the behalf of another but we did not use this technique in our
prototype.

The RTPS protocol supports reliable delivery of DDS samples through
acknowledgments and retries.  In HodorDDS. when a negative
acknowledgment (NACK) of a packet is received, the thread that
discovered the NACK (by polling the NIC) is responsible for sending a
retry of the lost packet.  This will often be the daemon described in
Sec.~\ref{sec:dds-daemon}, but it will sometimes be an ordinary
application thread that happens to poll when a NACK is waiting in the
device RX queue.

\subsection{Waiting and Eager Notification}
\label{sec:hodordds_EN}

When a DDS application wishes to suspend until some condition is met, it
can use a DDS \emph{waitset} to indicate a set of conditions for it to
resume execution.  Our prototype supports the use of waitsets to specify
that the owning thread should resume when a sample becomes available on
specified DDS topics.  Once a waitset has been properly initialized
inside the protected library, the owning thread uses a wrapper to
suspend outside of the library as described in
Sec.~\ref{sec:suspending}.  Other threads in the library check for
relevant waitsets when performing operations that may trigger them.

Since DDS includes objects named \emph{writer} and \emph{reader}, these
terms are here used only in reference to these objects. We will use the
terms \emph{deliverer} and \emph{receiver} to refer to the roles of
threads executing in the protected library, regardless of the origin of
the sample being delivered.

In HodorDDS, data can become available for readers in two ways. First, a
writer thread may deliver data directly to a topic with readers located
on the same physical machine (note that if the topic has any remote
peers, the writer must also send the data over the network).  Second,
DDS samples may arrive via the network and be processed for delivery by
any local thread executing in the protected library (usually this will
be our \emph{trusted daemon} process, described in
Sec.~\ref{sec:dds-daemon}).  In either case, we employ \emph{eager
notification} to reduce overall data movement latency between deliverers
and receivers as Sec.~\ref{sec:eagernotification} describes.  Once a
sample is ready for delivery, the deliverer checks whether any wait
conditions are associated with the destination topic.  If so, the
deliverer will immediately wake the indicated sleepers (using the
mechanism described in Sec.~\ref{sec:suspending}) so that they can begin
to enter the library while delivery is being completed.  The delivering
thread then increments the sample reference count, and provides each
subscribed reader with a \emph{message receipt} that identifies both the
topic and the location of the buffer containing the sample data.  These
operations take modest but linear time with respect to the number of
receivers and are not impacted by other parameters such as the message
size.  While this eager notification can, in principle, cause the
receiver thread to enter the library prematurely and spin, the fact that
delivering threads are never interrupted while in the protected library
means that the spin is guaranteed to be bounded (and thus complies with
the bounded-time rule).  Consequently, our prototype's latency was
reduced by as much as 12\% for 64\,B samples at low sample rates.

\subsection{Usage of Permanently Mapped Buffers}

HodorDDS assumes the availability of permanently mappable buffers
(Sec.~\ref{sec:permbuf}) to transport DDS sample data between
applications and the protected library.  We use a simple allocator to
allow the protected library and the application to reserve sample-sized
blocks of the buffer for communication in either direction and to free
them once they are not needed.  Each block of the buffer is also marked
with a status that indicates whether the sample it contains is ready to
be read.

When a sample is sent, the permanent buffer block containing the sample
is temporarily attached to other data structures in the protected heap,
making it reachable by peers. This allows local communication with only
a single copy step in which the receiver copies a sample directly from
another application's permanent buffer to its own.  After a sample has
been read by all local peers, and once it has been positively
acknowledged by all networked peers, the buffer block can be freed back
into the permanent buffer it originated from, where it can be
reallocated by either the application or the protected library to send
or receive additional samples.

The Hodor Linux kernel currently lacks the new features described in
Sec.~\ref{sec:permbuf}.  To emulate permanent buffers, our HodorDDS
library maps a large region shared by all processes (including the
daemon).  This region is not protected with MPK or by any other means.
Individual ranges of this region are assigned to each application
process when it enters the library for the first time.  In accordance
with our proposed protections, applications running outside of the
protected library only access their assigned range within the region,
and library code ensures that pointers passed by applications fall
within those ranges.  We believe this emulation faithfully captures the
capabilities and cost of actual permanent buffers.

\subsection{HodorDDS Daemon}
\label{sec:dds-daemon}

HodorDDS uses a trusted daemon process to respond to asynchronous
events, as Sec.~\ref{sec:async} describes.  More specifically, it relies
on the daemon for three things.  First, the RTPS protocol used by most
DDS implementations requires that a \emph{heartbeat} packet be sent at a
configurable interval by writer entities.  The HodorDDS daemon examines
the library state and sends heartbeat packets on the behalf of any
existing writers at the appropriate time intervals.  Second, when
HodorDDS is operating in polling mode, the daemon regularly checks
whether the NIC has received any packets and, if so, it processes them,
notifying relevant application threads via the futex mechanism described
in Sec.~\ref{sec:suspending}.  Third, when HodorDDS is operating in
interrupt mode, the daemon wakes promptly in response to newly arrived
network packets and notifies relevant application threads in this case
also.

Although the daemon reliably bounds the latency of packet processing in
both polling and interrupt mode, any thread entering the protected
library may also poll for packets and process them for itself or on the
behalf of any other participant, even when the daemon is running in
interrupt mode.

To implement the interrupt-driven mode outlined in Sec.~\ref{sec:async},
we take advantage of existing DPDK functionality that allows a DPDK
application thread to suspend (via a wrapper to the
\pcode{epoll\_pwait()} syscall) until the DPDK-managed NIC has received
a packet.  If the daemon is suspended using this mechanism, a hardware
interrupt from the DPDK-managed NIC will cause the OS kernel to wake the
daemon.  The daemon then retrieves and processes any packets waiting in
the device RX queue, and notifies the appropriate recipients using the
futex mechanism before completing delivery.  While running, the daemon
measures the packet rate and switches dynamically between interrupt and
polling modes depending on whether the average inter-arrival time is
long enough to accommodate a non-trivial sleep in interrupt mode.

\section{Performance Evaluation}
\label{sec:performance}

\subsection{Methodology}
\subsubsection{Test Configuration}
\label{sec:configuration}

Our tests were conducted on a pair of single-socket machines equipped
with Intel Xeon Silver 4208 CPUs with 8 cores/16 hyperthreads running at
2.10~GHz. The `relay' machine was equipped with 16 GB DDR4 RAM and the
`driver' machine was equipped with 32~GB (our tests used far less than
these amounts and the difference is not significant).  The machines were
connected via 10 gigabit ethernet using a pair of Intel X550-T2 NICs.
After finding that CPU throttling was causing erratic fluctuations in
latency (due to low CPU load between DDS samples), we disabled frequency
scaling, including Turbo Boost and hardware-controlled performance
states (HWP)\@.  We note, however, that HodorDDS performance was better
than the competitor system by a similar factor even with throttling
enabled: it was disabled only to rule out other possible sources of
variance, and to enhance the explainability of our results.

We did not implement the nonpreemption kernel features described in
Secs.~\ref{sec:hodor} and~\ref{sec:design}.  To minimize the impact of
preemption on our performance results (particularly given our use of
spin locks), we pinned each application or daemon thread to a separate
hardware context and avoided running other applications on the machine
concurrently.

We compiled the Hodor Linux kernel and HodorDDS using GCC version 11.40.
We compiled Linux with its default compiler flags.  We compiled HodorDDS
at optimization level \texttt{-O3}.  We built our competitor system,
FastDDS 2.14, with default ``Release'' parameters.  FastDDS and HodorDDS
were both configured to send one heartbeat packet per second,%
\footnote{This value resulted in low CPU utilization and was sufficient
    for the very low rate of packet loss in our tests.  On a less
    reliable network, more frequent heartbeats can partially compensate
    for packet loss by triggering retries.}
and DDS writers were configured with the following QoS settings:
RELIABLE\=RELIA\-BIL\-ITY\=QOS, KEEP\=ALL\=HISTORY\=QOS, and
TRANSIENT\=LOCAL\=DURA\-BIL\-ITY\=QOS.

\subsubsection{Test Application and DDS Usage}

We used the same C++ application code for HodorDDS and FastDDS: the only
differences involved the linking and initialization
of the appropriate DDS library.%
\footnote{In addition to the standard API, FastDDS supports listener
    callback functions. We did not use these in our tests.}
Our benchmark application comprises three single-threaded processes, and
is designed to measure the round trip time between two networked
machines.  We refer to these three processes as the \emph{driver}, the
\emph{relay}, and the \emph{listener}.  The \emph{driver} process (on
machine~1) sends DDS samples over the network to the \emph{relay}
process (on machine~2), which then relays them over the network again to
the \emph{listener} process (on machine~1).

The \emph{driver} process generates and sends samples at regular
intervals.  It records the time at which the first sample was sent and
then calculates the time at which the next sample should be sent. If the
next sample is not yet due to be sent, the \emph{driver} sleeps using
{\tt clock\_nanosleep()} until it is time to send the next sample;
otherwise, it sends the sample immediately. The deadline for the
previous sample is then used as the base to compute the deadline for the
next sample. This continues until the number of samples specified for
the current test have been sent.

Each test is conducted with a specific sample size.  The driver-relay
connection and the relay-listener connection use two separate DDS
topics, each initialized with the same sample size and QoS settings.

Aside from a timestamp inserted by the \emph{driver} process, the DDS
samples contain only zeros as filler data.  When the \emph{listener}
receives a sample, it compares the contained timestamp against the
current time to compute sample latency.

\subsubsection{Data Collection and Plotting}
\label{sec:datacollection}

We observed significant tail latency in FastDDS; Approximately 5\% of
FastDDS samples were two orders of magnitude slower than the average,
which would have significantly skewed the results in favor of HodorDDS.
Because we use FastDDS as representative of the kernel-based approach,
and because we do not believe that the high tail latency is inherent to
this approach, we have chosen to ignore the worst 10\% of FastDDS
samples when computing the average latency of FastDDS in all tests.

In the 1\,MB test (shown in Fig.~\ref{fig:plot:array1m}), we observed
extremely wide variances in the latency of FastDDS (even discarding the
worst 10\% as described above).  For this reason we use error bars to
show the range of results out of 5 trials for each data point, with the
point itself representing the best achieved time.  In the same test,
HodorDDS shows a maximum variance of less than 1 percent; we have
omitted error bars for HodorDDS because the upper and lower bars would
be indistinguishable.  We believe that the abnormally wide variation in
latency in this test was due to occasional packet loss when the sample
rate was sufficiently high to prevent the application from successfully
catching up.  As with the tail latency described above, we do not
believe that this is inherent to the FastDDS design approach.  We
therefore compare our results against the best of the five FastDDS
trials, but we report the range of all five trials for completeness.

To measure CPU utilization, each process running on the machine hosting
the \emph{relay} process used the {\tt times()} syscall to retrieve the
amount of time spent executing in user mode and kernel mode.  After the
expected number of samples for the test had been processed, each process
then used {\tt times()} a second time and computed its own CPU
utilization by summing the elapsed user and system execution time and
then dividing that by the wall clock time multiplied by 16 (as there are
16~hardware threads).  That is, the CPU utilization values are with
respect to all 16 hardware threads of the machine hosting the
\emph{relay} process.  With 16 hardware threads, a single thread can use
at most 6.25\%. As both HodorDDS and FastDDS use one application thread
and one daemon thread, both will use at most 12.5\% in these plots.

\subsection{Results}

Fig.~\ref{fig:plot:array64b} shows the latency and CPU utilization of
sending 64\,B messages from one machine to the other and back at varying
message frequencies.  This sample size is considerably smaller than the
network's Maximum Transmission Unit (MTU), meaning that the overhead of
sending each packet is maximized relative to the overhead of payload
data movement.  HodorDDS maintains the lowest latency as well as the
lowest CPU utilization across all message frequencies.

\begin{figure*}
\hspace*{-1em}
\centering
\includegraphics[width=.95\textwidth]{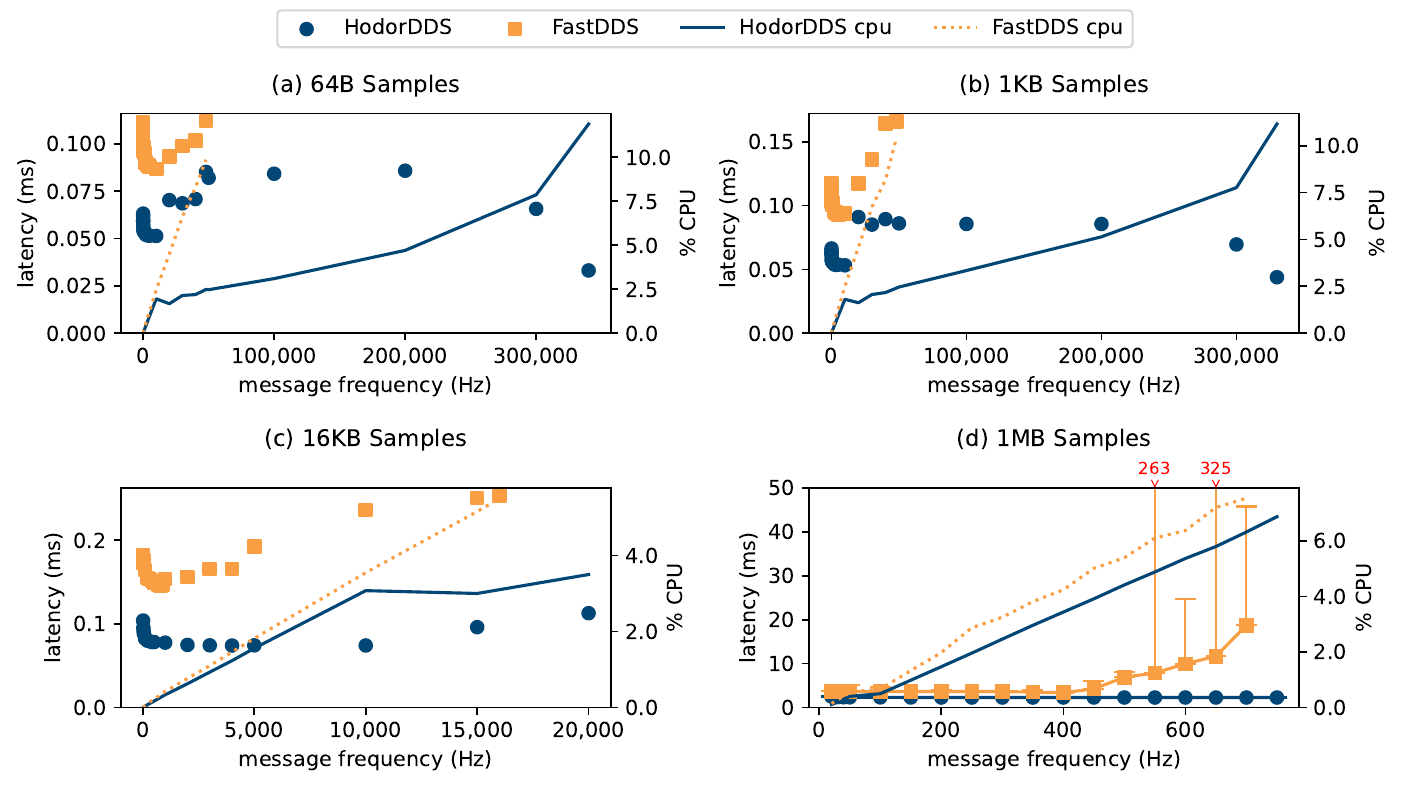}
  \Description{Four charts, one each for 64B, 1KB, 16KB, and 1MB
    samples, with latency and CPU utilization shown in each.}
  \caption{Message-passing latency (glyphs) and CPU utilization
      (curves) for HodorDDS and FastDDS.}
    \label{fig:plot}
    \extralabel{fig:plot:array64b}{(a)}
    \extralabel{fig:plot:array1k}{(b)}
    \extralabel{fig:plot:array16k}{(c)}
    \extralabel{fig:plot:array1m}{(d)}
\end{figure*}

Above 10,000~Hz, the amount of CPU time spent processing interrupts in
the kernel and changing privilege modes becomes too significant to allow
most samples to be processed before the next one arrives.  This is
therefore the point at which HodorDDS chooses to switch from interrupt
to polling mode.  FastDDS, by contrast, remains in interrupt mode.  It
begins to experience a very sharp increase in average latency, and is
never able to exceed about 48,000 samples per second.  HodorDDS,
however, is able to sustain some 340,000 samples per second at the 64\,B
size.  At this sample rate, the CPU utilization of the daemon is near
the theoretical limit for one hardware thread, meaning that it rarely or
never suspends.  Whenever the daemon has finished processing one sample,
there is usually another one waiting, so the daemon continues
running. This leads to even lower latency than the interrupt-driven mode
could achieve.

Fig.~\ref{fig:plot:array1k} shows analogous latency and utilization
results with 1\,KB messages.  At this size the payload is close to the
size of the MTU, but sample fragmentation is not yet needed.  These
results closely resemble those shown in Fig.~\ref{fig:plot:array64b},
with slight differences in overall shape as the overhead of copying the
payload is more significant in this test.  As in the prior figure,
FastDDS is unable to sustain a rate higher than 48,000 samples per
second.

Fig.~\ref{fig:plot:array16k} shows analogous latency and utilization
results with 16\,KB messages.  At this size, HodorDDS is forced to
divide each sample into multiple fragments due to the MTU limit.
FastDDS instead relies on the OS kernel to fragment at the MTU level; it
does its own chunking only when messages exceed 64\,KB in size.  The
range of frequencies shown in this test is more limited, but the change
to HodorDDS polling mode can still be observed above 10,000~Hz.

Fig.~\ref{fig:plot:array1m} shows analogous latency and utilization
results with 1\,MB messages.  Neither system was able to sustain a
sample rate at which interrupts become less efficient than polling in
this test, so the HodorDDS daemon is running in interrupt-driven mode
for this entire plot.  As explained in Sec.~\ref{sec:datacollection},
FastDDS exhibited wide latency variance for several trials shown here
(especially at 550 and 650 samples per second).  We report 5 trials per
data point for completeness, but plot the curve through the best of
these 5 trials.  FastDDS shows a gradual increase in latency throughout
this plot while HodorDDS remains more stable.  This stability stands in
contrast to the other plots and is due to the high degree of
fragmentation for 1\,MB samples.  A sample can only be forwarded by the
relay once all fragments have been received, which means that the
variance is reduced as long as each sample is fully processed before
reassembly of the next sample begins.  If the \emph{driver} sends more
than 700 samples per second, however, samples begin to overlap and the
throughput drops.

We also ran several tests to quantify the benefits of eager
notification.  For 64\,B messages at 10 samples per second, HodorDDS
achieved a round trip latency of about 62\,$\mu$s, as shown in
Fig.~\ref{fig:plot:array64b}.  Without eager notification, the resulting
latency was about 70\,$\mu$s, or about 14\% higher.  At 100 samples per
second, the advantage of eager notification decreases significantly,
achieving a latency of 57.1\,$\mu$s versus 57.7\,$\mu$s, and the
difference for even higher frequencies (or for larger samples) becomes
essentially negligible.  In our prototype, the amount of work done by a
thread after notifying peers is rather small.  If the deliverer
encounters cache misses on HodorDDS internal structures, however---as is
likely when there is a longer interval between samples---delivery takes
long enough to see some benefit from eager notification.  The gains
would likely be noticeable even at higher sample rates if the deliverer
were able to perform more useful work after waking the receivers---e.g.,
by beginning to copy data into the receiver's permanent buffer.

\section{Conclusions and Future Work}
\label{sec:future}

In this paper we have identified and resolved several previously
unaddressed obstacles to using protected libraries for shared
kernel-bypass services.  Our work makes it possible, for the first time,
for mutually untrusting applications to interact safely and
synchronously in user space, with direct but protected access to
memory-mapped devices.  Our work also highlights several new subtle
safety and performance considerations along with opportunities for
performance gains.  First, we show how synchronous interaction can be
achieved despite the requirement that protected library operations
complete in modest, bounded time.  Second, we show how to leverage the
bounded time requirement to achieve lower, more stable operation
latency.  Third, we identify a buffer unmapping vulnerability in prior
work on protected libraries, and provide a solution that introduces no
run-time overhead and in fact enables additional performance
optimizations (i.e., fewer copies) for synchronous interaction.  Fourth,
we consider the need to respond efficiently to asynchronous events in
protected libraries. To address this need, we introduce a \emph{trusted
daemon} to field asynchronous events and show how this daemon can
leverage shared state and kernel bypass to divide work with application
threads in pursuit of high throughput and low, stable latency.

Additionally, we have presented essential guidelines for protected
library developers to facilitate synchronous interaction and to maintain
safety while maximizing performance.  We have also made recommendations
for the designers of protected library operating systems, to address
deficiencies in prior work that impact the safety of both synchronous
interactions and data movement.

Finally, to validate the above contributions, we have prototyped a
protected library for networked IPC using the DDS middleware standard.
Compared to a traditional DDS implementation, HodorDDS improves latency
and throughput several-fold, with lower variance and computational load.

We believe that there are other promising opportunities for protected
libraries to manage shared resources such as data acquisition services,
storage, high-bandwidth output devices, and computational accelerators.
Exploration of these uses may yield further opportunities to improve the
generality and performance of protected libraries.

We envision several directions for future work.  For example, although
we currently redirect hardware interrupts exclusively to the daemon
process, it might be better to direct certain interrupts directly to
specific applications. The performance tradeoffs relative to our
approach are non-trivial. An untrusted application needs additional time
to enter the library compared to the trusted daemon; the daemon can
respond to interrupts more quickly. On the other hand, in cases where
the application must take some action itself, the improved notification
latency might be beneficial.

Bounded wait and condition synchronization might also be improved.  One
might, for example, partition a protected library into ``critical'' and
``non-critical'' segments.  The kernel would enforce the bounded wait
requirement only for threads in the critical segment.  Library designers
could then use the non-critical segment (with free, subroutine-call
transitions in and out of the critical segment) for operations
(including futex wait) that could safely be preempted or, in the case of
error, terminated.  Alternatively, a mechanism such as Solaris
\pcode{schedctl}~\cite{dice-schedctl} or Symunix
\pcode{tempnopreempt}~\cite{edler-uwus-1988} could allow a library
thread to bracket critical sections dynamically.

In summary, we see protected libraries as a way to finally realize the
potential of the microkernel concept, with a wealth of opportunities for
future research and development.

\begin{acks}
This work was supported in part by NSF grants CNS-1900803 and
CNS-1955498, and by a Google Faculty Research Award.
\end{acks}

\bibliographystyle{ACM-Reference-Format}
\bibliography{main}

\end{document}